\documentclass[10pt,a4paper]{IEEEtran}

\usepackage{color}
\usepackage{ifthen}
\usepackage{times}
\usepackage{cite}

  \usepackage[pdftex]{graphicx}
  \DeclareGraphicsExtensions{.pdf,.jpeg,.png}
\def\paragraph#1
{
  \smallskip
  \noindent{\bf #1}\hspace{.5em}
}

\def\USCORE{\_}

\newcommand{\Draft}{0}   % 0 = final, anything else is rough draft

\ifthenelse{\Draft = 1}
{
  \newcommand{\fixMe}[2][]{
    \typeout{***** ERROR: fixMe still in final version *****}
  }
}
{
  \newcommand{\fixMe}[2][] {[{\bf #1}] {\bf \marginpar{\large FIX}} {\em #2}}

}

\begin{document}

\title{Simultaneous Event Execution in\\Heterogeneous Wireless Sensor
  Networks}
\author{
    Tobias Baumgartner, S{\'a}ndor P. Fekete, Winfried Hellmann,\\
    and Alexander Kr{\"o}ller\\
    IBR, Algorithms Group\\
    Braunschweig University of Technology\\
    M{\"u}hlenpfordtstra{\ss}e 23\\
    38106 Braunschweig, Germany\\
    Email: \{t.baumgartner,s.fekete,w.hellmann,a.kroeller\}@tu-bs.de
}
\date{}

\maketitle

\begin{abstract}
  We present a synchronization algorithm to let nodes
  in a sensor network simultaneously execute a task at a given point in
  time. In contrast to other time synchronization algorithms we do not
  provide a global time basis that is shared on all nodes. Instead,
  any node in the network can spontaneously initiate a process that
  allows the simultaneous execution of arbitrary tasks. We show that
  our approach is beneficial in scenarios where a global time is not
  needed, as it requires little communication compared with other time
  synchronization algorithms.

  We also show that our algorithm works in heterogeneous systems
  where the hardware provides highly varying clock
  accuracy. Moreover, heterogeneity does not only affect the
  hardware, but also the communication channels. We deal with
  different connection types---from highly unreliable and
  fluctuating wireless channels to reliable and fast wired
  connections.
\end{abstract}


\section{Introduction}
\label{sec:intro}

When dealing with algorithms for sensor networks, time synchronization
is an important factor. For many kinds of application areas it is
essential to have a common global time basis. This can be required due
to the need for putting certain events within the network in a
chronological order, to execute tasks synchronously, or for clock
synchronization in time division multiple access (TDMA) based media
access layer (MAC) protocols.

While there are already standards for ordinary networks available, such as the Network Time Protocol (NTP) or Precision Time Protocol (PTP), it is still a challenging task for sensor networks. This is due to several reasons. First, we mainly have wireless communication in sensor networks, and thus a higher amount of unreliability and fluctuation. Second, we must deal with energy constraints. The nodes may be operated with batteries, and hence they are not able to exchange messages frequently over a long period of time. Third, because of low hardware costs the nodes may have a high clock drift---that is, clocks run at a slightly different speed, and so they drift apart even if started at exactly the same time.

There have already been many algorithms developed that deal with these
issues. However, there are application scenarios where a consequent
time synchronization, with all nodes sharing the same time basis is
not needed and would produce too much overhead. For example, if
the nodes only need to start a common task at a certain point in time,
but do not need any common time basis apart from that, it is possible
to use much simpler algorithms. So the network can stay unsynchronized
most of the time, but only collaborate shortly before
the designated event.

An application for such a system is collaborative sensing of highly dynamic effects. For instance, to locate the source of an audio signal, it is necessary to collect synchronized readings from the sensor network. An initiator node would then start the process so that every node measures the local volume at the very same point in time. The resulting map can then be used to determine the source of the signal. It is obviously crucial that all nodes collect their data at a synchronized point, yet it is not necessary to keep the network synchronous at all times. Another example is simultaneous output, e.g., the playback of music by a sensor network where each node is equipped with a speaker. Here, the sensors must start the playback at a particular time, but any synchronization before the event is not needed.

However, even such a simplification of the requirements may lead to a
goal that is hard to achieve. This is especially the case when dealing
with heterogeneity in sensor networks. There may be very different
kinds of hardware platforms used. Thus, there is a high variance in
clock drift among the used architectures. Some platforms may be
equipped with very accurate clocks, whereas others provide only very
course ones. Moreover, the nodes may communicate over different
communication channels---from 2.4 GHz IEEE 802.15.4 over the 868 MHz
band to wired Ethernet or even SPI or RS232.

This paper presents an algorithm that deals with the above described
issues. First, our algorithm provides a spontaneous, on-demand
synchronization mechanism. That means, at an arbitrary point in time,
any node in the sensor network is able to act as a master and initiate
a synchronization process. The exclusive goal of this process is that
all participating nodes can start a task at the same point in time,
relative to a start signal from the master. Hence, the nodes do not
need a global time basis, and thus messages required for continuous
synchronization are completely avoided.

We also address interference issues in our algorithm. To avoid all
nodes sending replies to the master at the same time, causing
interference in a broadcast medium, the algorithm needs only two
participants: one master and one slave. The other nodes in
communication range merely overhear the communication without sending
any message, but are still able to join in the synchronization event
execution. This leads to a considerable reduction of energy
consumption and interference.

Second, our algorithm can deal with both heterogeneous networks and
hierarchical structures. The former enables the use of different
hardware platforms, whereby especially various clock accuracies were
taken into account. The latter enables the possibility of building a
layered structure of participants. This can be, in the simplest
topology, a wireless multihop connection to nodes that can not be
directly reached by the master. But it can also be a wired connection
between two nodes of completely different hardware platforms.

We show that our approach works accurately with different hardware
platforms. Therefore we chose two architectures with different
capabilities: One more powerful sensor node that runs quite
complicated algorithms, and a very restricted one with supremely
limited code space and an inaccurate clock. There are also
different communication channels available. Some of the nodes
communicate via their radio, whereas others are connected via a wired
connection and communicate over SPI.

The next section describes the related work that has already been done
in this context. Section~\ref{sec:algorithm} presents our algorithm in
detail, followed by the experimental
results. Section~\ref{sec:conclusion} concludes this paper.

%%% Local Variables: 
%%% mode: latex
%%% TeX-master: "wsn09-flashmob"
%%% End: 

\section{Related Work}

Time synchronization has been an issue in distributed systems for long time now.
Thus an abundance of algorithms is already available~\cite{DBLP:journals/adhoc/SundararamanBK05,1285634,roe05}. Many of them are unsuitable for wireless sensor networks, as they rely on stable communication,
complex calculations, or high clock accuracy.

The remaining algorithms can be categorized roughly by the following properties:
\begin{itemize}
\item {\em Relative synchronization} calculates the drift between the clocks of two nodes,
  instead of adjusting them to a central time source.
\item {\em Passive nodes} perform synchronization without ever communicating themselves, they merely listen to the communicating of others.
\item {\em Event-triggered synchronization} proceeds only on demand, while {\em continuous
  synchronization} keeps the clocks aligned at all times.
\end{itemize}

The Reference-Broadcast Synchronization (RBS)~\cite{elson_fine-grained_2002} is a
very common approach. It leads to accurate results and can be event
based, but it does not allow for passive nodes. Every pair of nodes has to exchange
messages containing their local timestamps at the reception of broadcast signal
in order to synchronize. This leads to a high communication rate during the
synchronization process, thus increased power consumption and risk of
collisions.

An improvement concerning performance is the Timing-sync Protocol for Sensor
Networks (TPSN)~\cite{ganeriwal_timing-sync_2003}. It builds a hierarchical
structure based on distances to one or more root nodes. Afterwards,
every node at distance $d$ to the root synchronizes with a node at distance
$d-1$.

%% unnecessary?
% Very similar works the Lightweight Time Synchronization~\cite{941353}, that
% uses a spanning tree instead of circles.

Our algorithm is neither similar to RBS nor TPSN. It employs a broadcast
element like RBS and a (dynamically changing) hierarchical master-slave structure
as in TPSN. Additionally it allows for passive slaves within communication range of the root, which are synchronized
without sending any message.

Another active synchronization algorithm is TinySync~\cite{1240228}. It
calculates clock drift by exchanging time stamps of two nodes and resolving 
the linear correlation. Though it allows two nodes to run synchronized
for a longer period, it has to perform complex computations and requires storage of many
data points. These two disadvantages make it inappropriate for systems
with less computing and storage capacity.

TSync~\cite{980173} allows for passive slaves. It uses one or more reference nodes to broadcast a certain beacon and the local timestamp, when sending it. A selected child node sends its own local timestamp at reception and the current one back. Therefore it is possible for the reference node to calculate the propagation time and the clock offset of the child. Indeed this algorithm is very similar to our approach. Unlike our approach, it does not calculate the clock drift and therefore has to be run periodically.

We already presented a preceding model of our algorithm in
\cite{flashmob_portalacm}. While the fundamental design
of the algorithm has not changed, we optimized the synchronization
process and obtained noticeably better results than in the previous
work. The enhancements concerned particularly the capability of
dealing with an implementation in application layer, but also a better
adaptation to heterogeneous networks.

% \begin{itemize}
% \item Standard time synchronization algorithms
% \item Are there simple ones like ours?
% \end{itemize}

%%% Local Variables: 
%%% mode: latex
%%% TeX-master: "wsn09-flashmob"
%%% End:

\section{Algorithm}
\label{sec:algorithm}

In contrast to other time synchronization algorithms, we do not
provide a global time basis on all nodes. Instead, our algorithm
allows any node in the network to spontaneously start a synchronous
execution of arbitrary tasks. Following this approach, there is no
need for sending periodic messages, and thus we achieve a remarkable
reduction of energy consumption in scenarios where a global time is
not needed.

We assume that any node in the network is allowed to initiate a
synchronization process. If so, the self-proclaimed master selects an
arbitrary node in communication range to start the
synchronization. The slave then obtains two values based on the
message exchange: propagation time of the message, and clock drift
with respect to the master's clock.

While the master and the slave calculate the required values, every
node in communication range of the master can determine the same
values for itself only by listening to the master's packets.

After that, each node then does the same procedure for its
unsynchronized neighbors, which are nodes outside the communication
range of the master. Finally, each node participating in the
synchronization process knows both the message propagation time and
clock drift to its predecessor. Then, the master broadcasts a message
with the given time span after which the nodes shall start their
execution. Each of these nodes is then able to calculate its local
waiting time. In the following, the different phases are described in
detail.

\subsection{Active Synchronization}

The active synchronization is done between a master and a slave,
whereby the slave is only supposed to calculate its own clock drift
relative to the master, and to obtain the propagation time of a
message.

\begin{figure}
  \centering
  \includegraphics[height=2.5in]{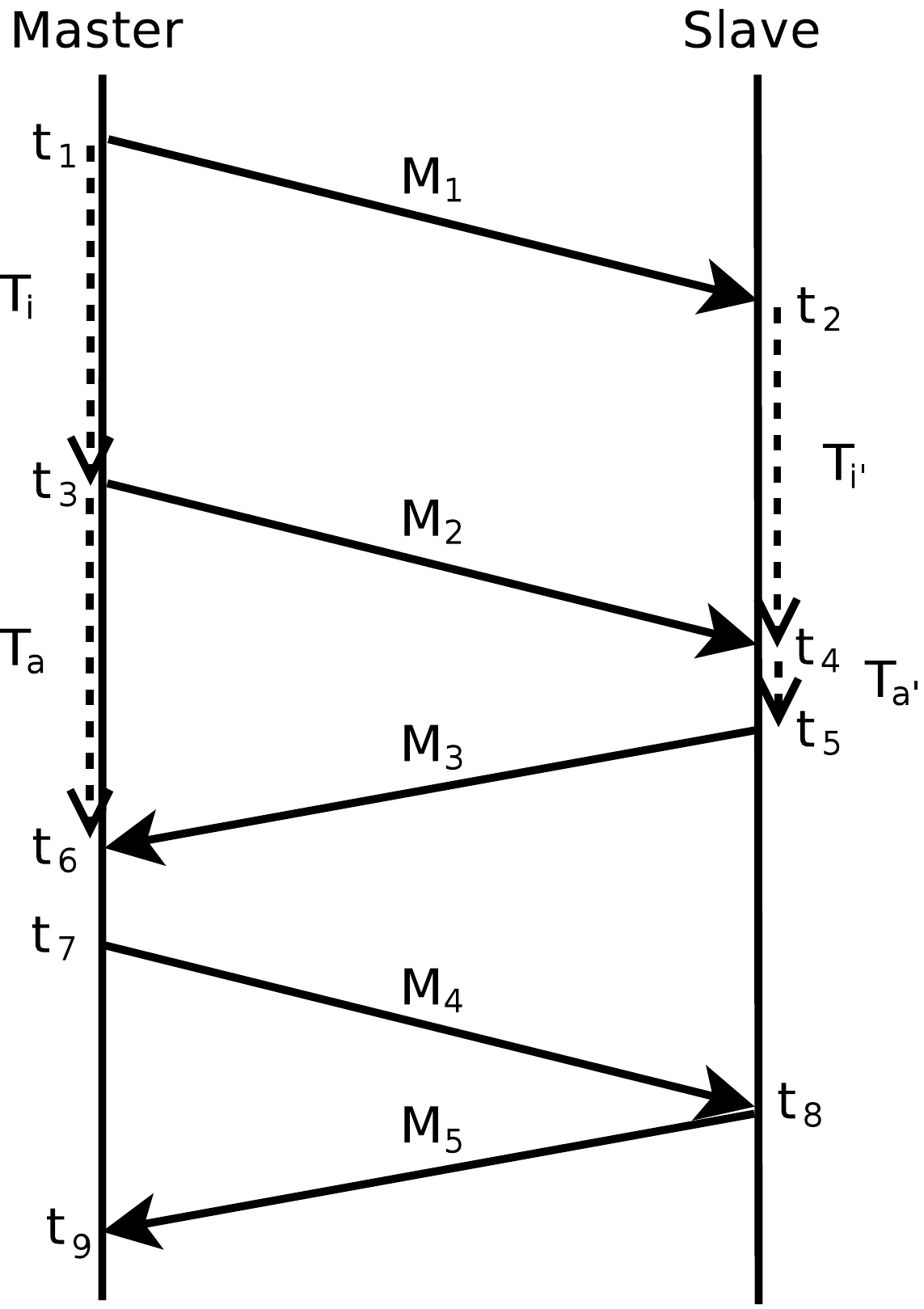}
  \caption{Active synchronization between master and slave.}
  \label{sec:alg:active_sync}
\end{figure}

The whole process needs only five messages, as shown in
Fig.~\ref{sec:alg:active_sync}. The master sends $M_1$ and
$M_2$ at $t_1$ and $t_3$, respectively. He then stores the time
interval $T_i$ between these events. The slave in turn takes the time
interval $T_{i'}$ of the reception of the messages, $t_2$ and $t_4$.

When receiving $M_2$ at $t_4$, the slave replies to the master with
$M_3$, and stores the time interval $T_{a'}$ between reception and
sending. The master in turn measures $T_{a}$, which is $T_{a'}$ plus
the propagation times of $M_2$ and $M_3$. $T_{a}$ is then sent back to
the active slave with $M_4$, which is then able to calculate the
propagation time of a message---finally, the active slave broadcasts
the propagation time for potential passive slaves (see next
subsection).

\subsection{Passive Synchronization}

When the master synchronizes with a slave, there may be other nodes in
communication range that can listen to the above described
conversation. Theses nodes are able to synchronize, too, if they are
at least able to hear the messages sent by the master. Doing so, they
do not need to send any message, and thus are able to conserve
energy. However, the propagation time between such a node and the
selected slave may be slightly different, but is still insignificantly
small with respect to other influencing factors such as computation
delays or interrupts delaying program execution.

\begin{figure}
  \centering
  \includegraphics[height=2.5in]{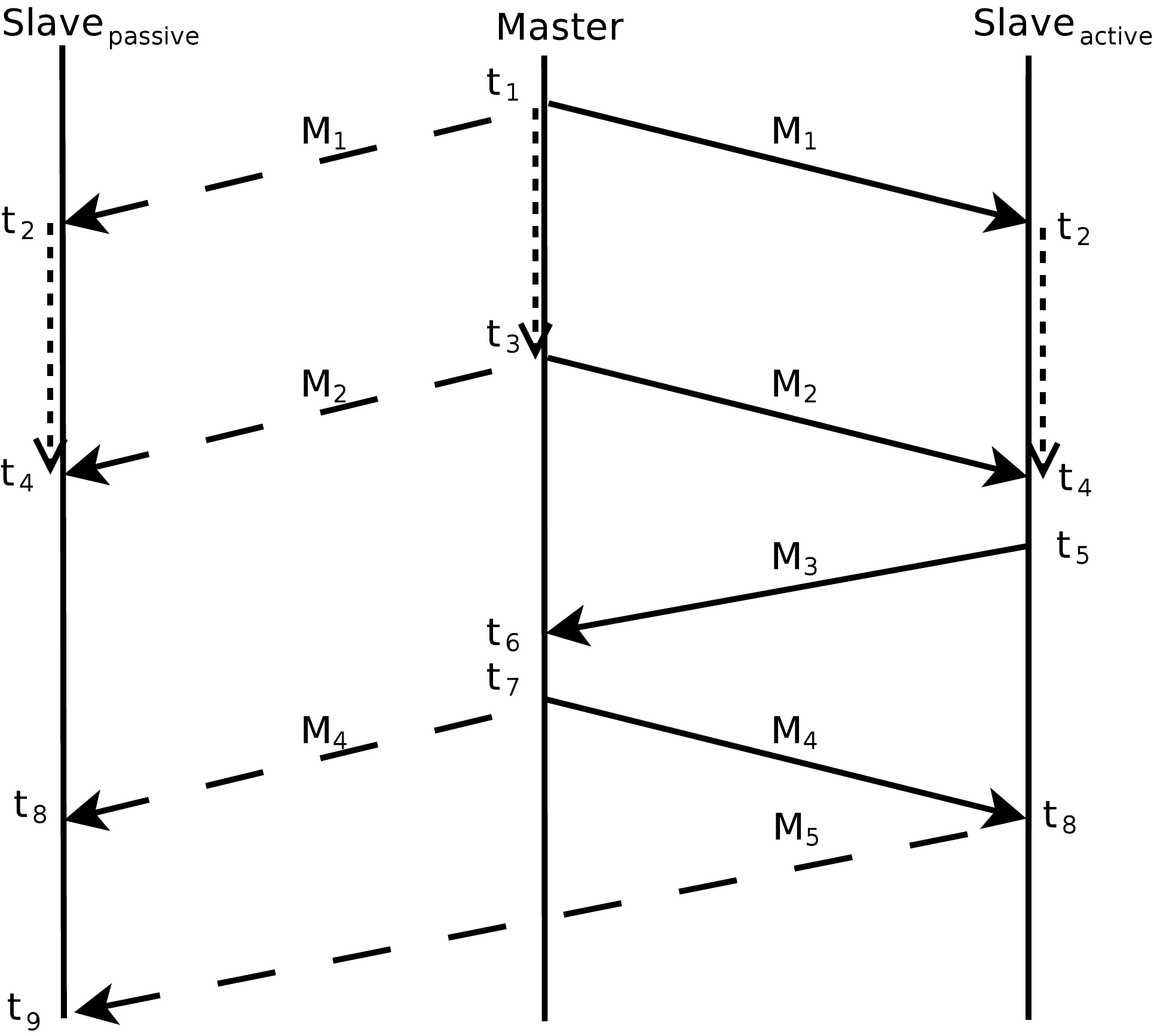}
  \caption{Passive synchronization of slaves in communication range of
    master.}
  \label{sec:alg:passive_sync}
\end{figure}

The process is shown in
Fig.~\ref{sec:alg:passive_sync}. Thereby the passive slave
receives messages $M_1$, $M_2$, and $M_4$ by the master. The slave can
then calculate its clock drift with the aid of the interval $t_4- t_2$
and the master's interval, which is contained in $M_4$. Moreover, the
passive slave receives the propagation time of a message in $M_5$,
which is sent by the active slave.

\subsection{Hierarchical Synchronization}

\begin{figure}
  \centering
  \includegraphics[height=1.5in]{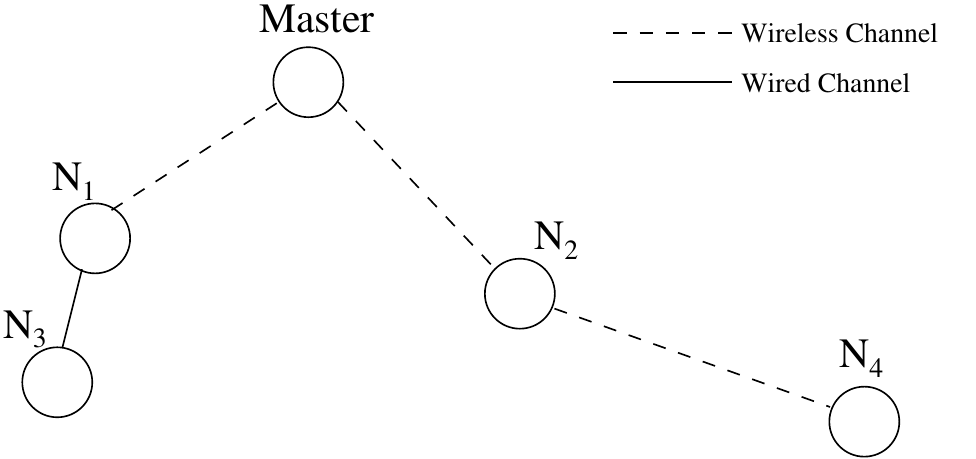}
  \caption{Hierarchical topology synchronized by master.}
  \label{sec:alg:hierachical}
\end{figure}

All nodes in communication range of the master are synchronized in the
previous steps---one by exchanging messages, the others by listening
to the conversation. However, there may be more nodes in the network
than the ones in communication range of the master. An example is
shown in Fig.~\ref{sec:alg:hierachical}.

There, the master synchronizes with nodes $N_1$ and $N_2$. After this
process has been finished, both of them synchronize with their
neighbors. Here, $N_1$ communicates to $N_3$ which is connected via a
wired channel, whereas $N_2$ communicates to $N_4$ over the radio.

The whole synchronization process works in exactly the same manner as
described previously. Hence, each node does a active synchronization
with one neighbor, and other neighbors can synchronize passively. This
way, the same code base can be used for all nodes, and also all
layers---independent of the hardware platform and communication
channel.

\subsection{Start Signal}

The final step of the event synchronization is sending the start signal. The master sends a message that contains the time interval after which the appropriate task should be executed by all the nodes in the network. Each node that receives this message calculates the time to wait based on the obtained values from the synchronization phase. This is done by subtracting the propagation time $T_\mathrm{propagation}$, and transforming the remaining time with the aid of $c_\mathrm{drift}$. The resulting time $T_\mathrm{wait}$ is then used to both register a timer for the global task execution, and forwarding to the neighboring nodes lower in the hierarchy. This process is repeated until all nodes in the network have set their timers.

%%% Local Variables: 
%%% mode: latex
%%% TeX-master: "wsn09-flashmob"
%%% End: 

\section{Experimental Results}
\label{sec:exp_results}

We implemented our algorithm on different hardware platforms being
connected via different communication channels. Thereby we had to deal
with several restrictions, from very limited memory to the lack of
access to the hardware, because the used OS did not provide such
functionality. Nevertheless, we got adequate results for our
application scenarios.

\subsection{Implementation}

We had two hardware platforms available for the implementation of our
algorithm. First, a tiny Atmel Atmega48~\cite{atmega48} with only 4 kB
of ROM and 512 bytes of RAM. It had also to run other applications, so
that there was only a very limited amount of memory available for the
synchronization process. Second, we used the iSense
platform~\cite{buschmann07isense} which is equipped with Jennic
microcontrollers~\cite{jennic}. These nodes provide a IEEE 802.15.4
compliant radio, and were already obtained with a running firmware
that was used for our implementation. It is an event-driven firmware,
and provides the registration of two types of callbacks: so-called
timeouts which run in interrupt context, and user tasks for
low-priority processing. User tasks can be only executed one after the
other---without the possibility of task switching, and without the
ability of dealing with different priorities. Timeouts, on the other
hand, can be executed while user tasks are running, because they run
in an interrupt service routine (ISR). Each time-critical part in the
synchronization process has been implemented using these
timeouts. However, since the platform does not allow that one ISR
interrupts another one, it is still possible that timeouts may be
delayed.

The Atmels were connected to iSense nodes via a wired connection, and
used SPI for communication. The iSense nodes in turn were able to
communicate over their radio. However, since we used the available
firmware we had no direct access to the MAC layer, and thus had to
implement the algorithm in the application layer---which obviously led
to a loss in accuracy.

Another issue were timer accuracies. On the Atmel platforms, we had to
use the internal oscillator, which led to timer interrupts of variable
length. The duration of such a timer event may vary from one call to
another---that means, running for a time of 2 milliseconds, for
example, may take +2$\mu$s at the first call, but -2$\mu$s at the
second call. We measured with an oscilloscope the exact period of such
events, and collected the minimal and maximal durations to obtain the
variance of the timer. To obviate influences during the test by other
tasks or interrupts on the Atmel, there was only the measurement
application running. The results for several periods are shown in
Table~\ref{sec:exp:atmel_clock}.

\begin{table}
  \caption{Timer variation on Atmel ATmega48 platform.}
  \center
  \begin{tabular}{ |c|c c c c| }
    \hline
    Period (ms) & Min (ms) & Max (ms) & Diff ($\mu$s) & Diff (\%) \\
    \hline
    0.2 &  0.2045 &  0.205 & 0.5 & 0.25 \\
    2   &  2.043  &  2.047 & 4   & 0.2 \\
    20  & 20.439  & 20.479 & 40  & 0.2 \\
    40  & 40.881  & 40.959 & 78  & 0.195 \\
    60  & 61.578  & 61.658 & 100 & 0.167 \\
    \hline
  \end{tabular}
  \label{sec:exp:atmel_clock}
\end{table}

Due to limitations of the oscilloscope, we were only able to measure
periods of up to 60ms---however, we see that even at 60ms, there is a
variation of up to 100$\mu$s. Projected on 500ms, this may result in a
variance of nearly 1ms.

Similar problems do also occur on the iSense platform. While iSense
nodes have a much more dependable clock than the Atmels, they run a
full firmware with message reception tasks, user tasks, and so
on. That means, even when we register a timer that is executed in
interrupt context, it may be delayed by other running interrupts in
the firmware. Table~\ref{sec:exp:isense_timer} shows results
measured with an oscilloscope.

\begin{table}
  \caption{Timer variation on iSense platform.}
  \center
  \begin{tabular}{ |c|c c c c| }
    \hline
    Timer (ms) & Min (ms) & Max (ms) & Diff ($\mu$s) & Diff (\%) \\
    \hline
    2  &  2.000 &  2.002 &  2 & 0.1  \\
    4  &  4.000 &  4.002 &  2 & 0.05 \\
    20 & 19.938 & 20.000 & 62 & 0.31 \\
    40 & 40.000 & 40.000 &  0 & 0.0  \\
    \hline
  \end{tabular}
  \label{sec:exp:isense_timer}
\end{table}

Whereas most of the timer events were very accurate, there are also
outliers due to firmware activity. An example can be seen at the 20ms
timeout, where a deviation of 62$\mu$s occurs. In addition, the iSense
firmware only offers timers with an accuracy of milliseconds, which
unfortunately makes an accurate synchronization in terms of
microseconds impossible.

\subsection{Experimental Setup}

\begin{figure}
  \centering
  \includegraphics[width=\columnwidth]{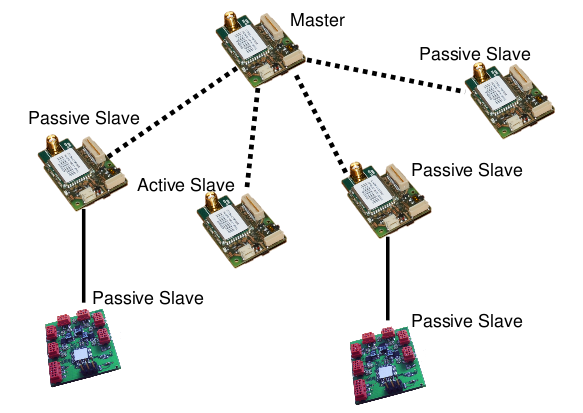}
  \caption{Experimental setup with six nodes building different kinds
    of slaves.}
  \label{sec:exp:exp_setup}
\end{figure}

The evaluation has been done with a small network of seven nodes---two
Atmels and five iSense nodes. One iSense node was set to the master,
and another three iSense nodes acted as passive slaves. In addition,
two Atmels were connected via wires to passive slaves. The setup is
shown in Fig.~\ref{sec:exp:exp_setup}.

All nodes---the four iSense nodes and the two Atmels---were connected
via wires to the interrupt lines of another microcontroller that
measured the exact points in time when the nodes fired their
events. Since the microcontroller was able to measure in terms of
microseconds, we were able to obtain correspondingly accurate
measurements.

\subsection{Results}

We ran several tests on our experimental setup. In each run, the
master synchronized the network spontaneously, and sent a start
message with different time intervals. We then measured the points in
time when the nodes executed their tasks.

Event though the algorithm would be able to deal with multiple
synchronization tasks at once---that is, a slave that synchronizes to
more than one master in parallel, or a slave that also acts as a
master---our experiments were only ran with one synchronization task
per experiment. The main reasons for this decision was a simplified
and more accurate measurement process, and a saving in code space,
which was especially important for the Atmels.

\begin{figure}
  \center
  \includegraphics[width=\columnwidth]{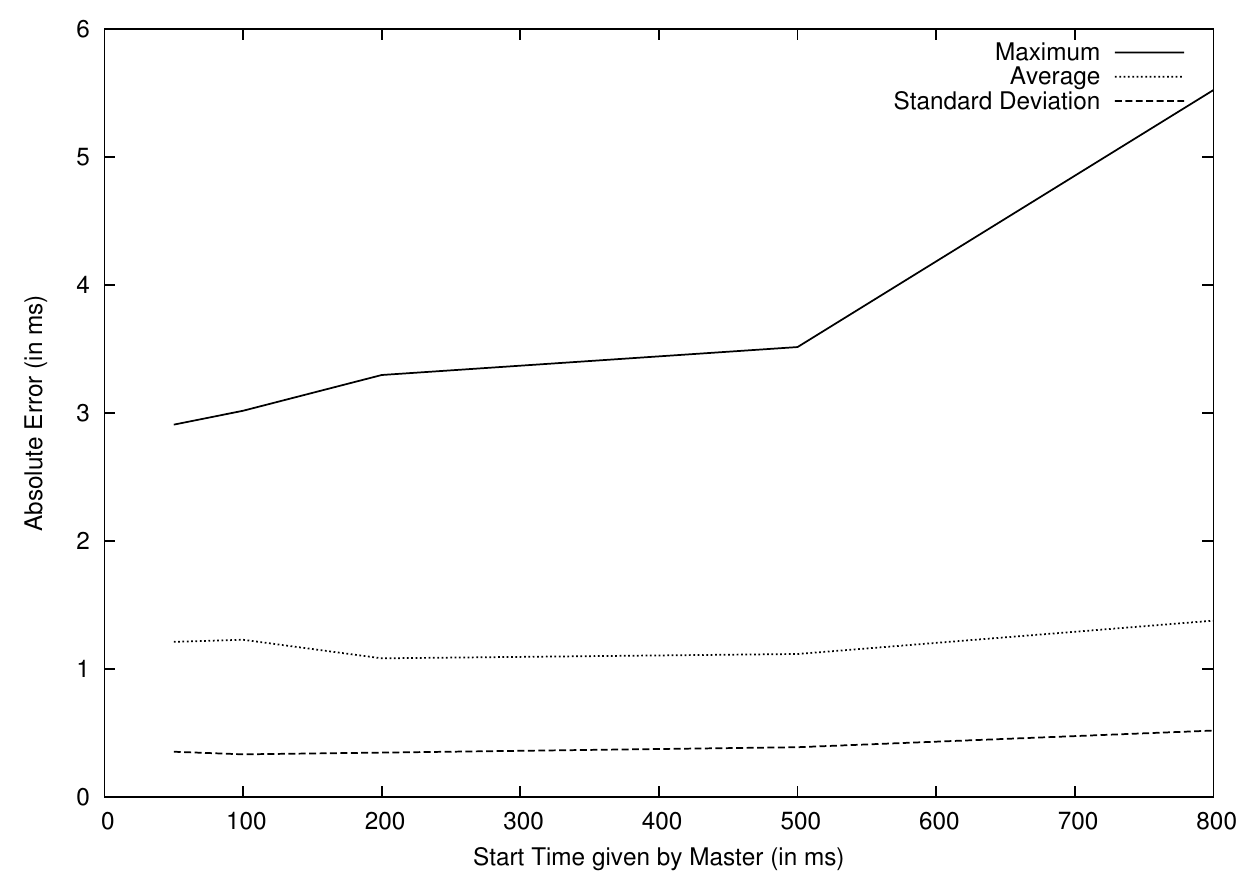}
  \caption{Absolute error in execution of events; start interval 50ms
    to 800ms.}
  \label{sec:exp:sync_variation}
\end{figure}

Fig.~\ref{sec:exp:sync_variation} shows the absolute errors
relative to the execution of the task at the master---as the average
over all six nodes, both iSense and Atmel ones. The start interval has
been increased consecutively from 50ms to 800ms.

It can be seen that the average error is at approximately 1ms,
regardless of the selected start interval. However, the greater the
start interval becomes, the more outliers occur, and thus the maximal
error increases to more than 5ms when selecting a starting time of
800ms. The maximal deviation in mainly caused by the very
inaccurate Atmel ATmegas. This difference in accuracy between the
iSense nodes and Atmels is shown in detail in
Table~\ref{sec:exp:errors}, which shows the error rates of all
nodes in our network when the starting interval is set to 500ms.

\begin{table}
  \caption{Absolute Errors in ms on nodes with a start time of 500ms}
  \center
  \begin{tabular}{ |l|c c c c| }
    \hline
    Node            & Min Err & Max Err & Avg Err & Stddev \\
    \hline
    Active Slave    & 0.004   & 1.455   & 0.640   & 0.389 \\
    Passive Slave 1 & 0.007   & 1.499   & 0.678   & 0.392 \\
    Passive Slave 2 & 0.006   & 1.463   & 0.734   & 0.375 \\
    Passive Slave 3 & 0.006   & 1.453   & 0.707   & 0.389 \\
    Atmel at PS 1   & 0.573   & 2.458   & 1.476   & 0.352 \\
    Atmel at PS 2   & 1.011   & 3.518   & 2.480   & 0.445 \\
    \hline
  \end{tabular}
  \label{sec:exp:errors}
\end{table}

All four iSense nodes show very similar results. The minimal error is
at only a few microseconds, whereas the maximal one is at most
1.5ms. In average, the deviation from the master is at around
700$\mu$s. In contrast to the iSense nodes, the Atmels show much worse
results. The deviations are between 500$\mu$s and 3.5ms, with an
average of 2ms. This is basically due to the inaccurate internal
oscillator, which makes it impossible to calculate reliable clock
drifts. However, for application areas where an event synchronization
of only a few milliseconds is adequate, such as a synchronous
playback of sound files or the concurrent sensing of a global event,
the number of only five messages that are sent over the radio
outperforms the inaccuracy.

\begin{figure}
  \centering
  \includegraphics[width=\columnwidth]{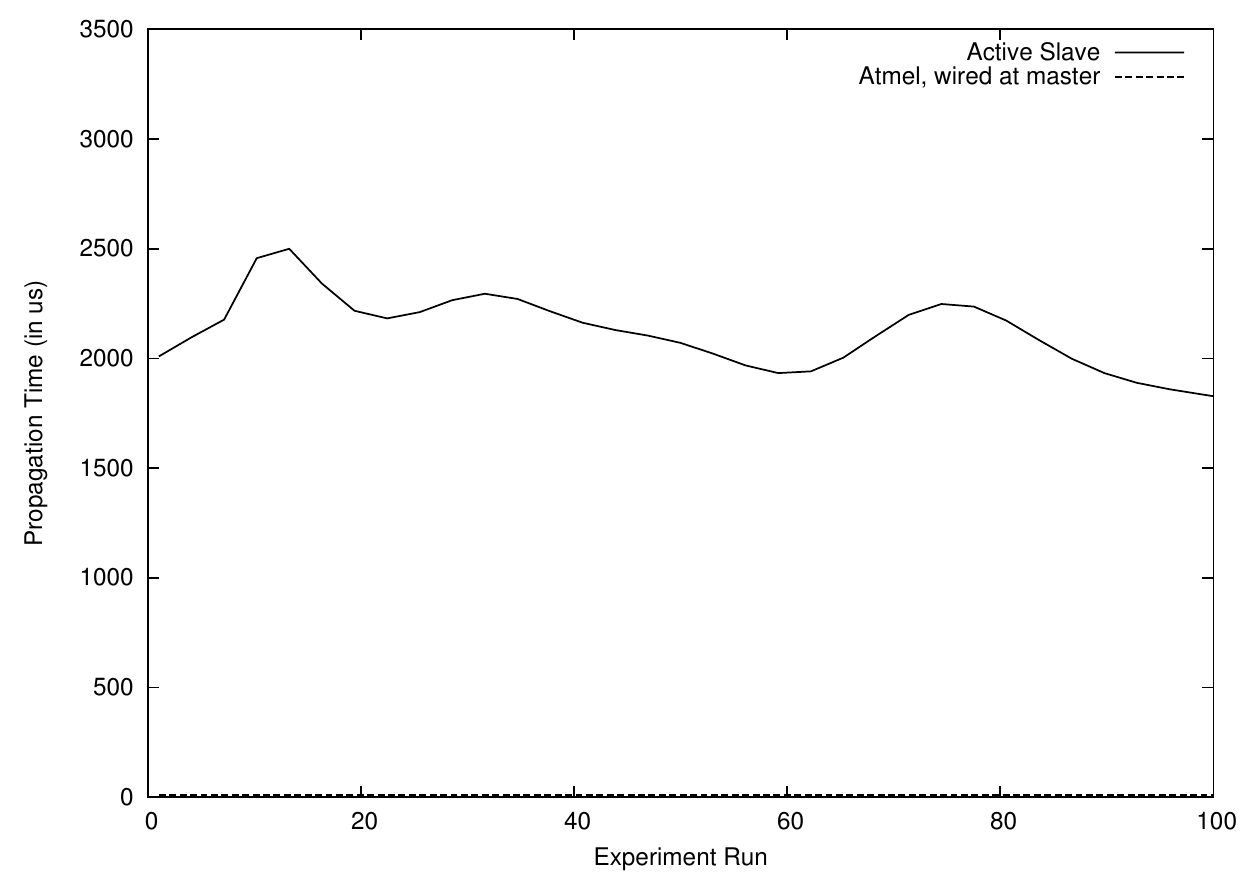}
  \caption{Propagation times over time.}
  \label{sec:exp:sync}
\end{figure}

Next, we also measured the propagation time of messages, both between
the master and the active slave, and an iSense node connected to an
Atmel via SPI. The result is shown in
Fig.~\ref{sec:exp:sync}.

The propagation time of the radio message varied between 2ms and
2.5ms, whereas the SPI communication was as expected constantly at
10us. The variance in wireless propagation time was mainly based on
the need for an application-layer implementation.
%However, even though the time varied
%by up to 500us, the results showed in Table~\ref{sec:exp:errors} were
%acceptable.

% \colorbox{red}{Weitere Tests}:
% \begin{itemize}
% \item Skript: x-Achse keine Zeit sondern Iterationen, Runden,
%   wasauchimmer
% \item Clock Drift auf verschiedenen Knoten zum Zeigen der
%   Heterogenitaet?
% \item Propagation Time? Varianz in Propagation Time auf iSense vs. konstanter
%   Wert mit Atmel/SPI.
% \item Die Abweichung bei verschiedenen Wartezeiten (100ms bis 2s)?
% \end{itemize}

%%% Local Variables: 
%%% mode: latex
%%% TeX-master: "wsn09-flashmob"
%%% End:

\section{Conclusion}
\label{sec:conclusion}

We presented an algorithm for the synchronous execution of 
tasks in sensor networks. In contrast to other well-known time
synchronization algorithms, we do not provide a global time basis on
the nodes. Instead, each node in the network is able to act
spontaneously as a master, and let the remaining nodes execute a task
at a certain point in the near future. The algorithm works in both hierarchical and
heterogeneous systems. The former can be used to synchronize the
network via multiple hops---if not all nodes are within
communication range of the master. The latter means that we dealt with
different kinds of hardware platforms and communication channels. Our
experimental setup consisted of both limited platforms with very
coarse clocks and nearly no code space, and ordinary sensor nodes with
IEEE 802.15.4 compliant radios. The nodes used also both wireless and
wired communication.

%%% Local Variables: 
%%% mode: latex
%%% TeX-master: "wsn09-flashmob"
%%% End:

\section*{Acknowledgment}

Tobias Baumgartner has been supported by the ICT Programme of the
European Union under contract number ICT-2008-224460 (WISEBED).

\bibliographystyle{IEEEtran}
\bibliography{paper}

\begin{thebibliography}{10}
\providecommand{\url}[1]{#1}
\csname url@rmstyle\endcsname
\providecommand{\newblock}{\relax}
\providecommand{\bibinfo}[2]{#2}
\providecommand\BIBentrySTDinterwordspacing{\spaceskip=0pt\relax}
\providecommand\BIBentryALTinterwordstretchfactor{4}
\providecommand\BIBentryALTinterwordspacing{\spaceskip=\fontdimen2\font plus
\BIBentryALTinterwordstretchfactor\fontdimen3\font minus
  \fontdimen4\font\relax}
\providecommand\BIBforeignlanguage[2]{{%
\expandafter\ifx\csname l@#1\endcsname\relax
\typeout{** WARNING: IEEEtran.bst: No hyphenation pattern has been}%
\typeout{** loaded for the language `#1'. Using the pattern for}%
\typeout{** the default language instead.}%
\else
\language=\csname l@#1\endcsname
\fi
#2}}

\bibitem{DBLP:journals/adhoc/SundararamanBK05}
B.~Sundararaman, U.~Buy, and A.~D. Kshemkalyani, ``Clock synchronization for
  wireless sensor networks: a survey,'' \emph{Ad Hoc Networks}, 2005.

\bibitem{1285634}
Y.~R. Faizulkhakov, ``Time synchronization methods for wireless sensor
  networks: A survey,'' \emph{Program. Comput. Softw.}, vol.~33, no.~4, pp.
  214--226, 2007.

\bibitem{roe05}
K.~R\"omer, ``Time synchronization and localization in sensor networks,'' Ph.D.
  dissertation, {ETH Z\"urich}, 2005, diss. ETH No. 16106.

\bibitem{elson_fine-grained_2002}
J.~Elson, L.~Girod, and D.~Estrin, ``Fine-grained network time synchronization
  using reference broadcasts,'' \emph{{ACM} {SIGOPS} Operating Systems Review},
  vol.~36, pp. 147--163, 2002.

\bibitem{ganeriwal_timing-sync_2003}
S.~Ganeriwal, R.~Kumar, and M.~B. Srivastava, ``Timing-sync protocol for sensor
  networks,'' in \emph{Proceedings of the 1st international conference on
  Embedded networked sensor systems}, 2003, pp. 138--149.

\bibitem{1240228}
S.~Yoon, C.~Veerarittiphan, and M.~L. Sichitiu, ``Tiny-sync: Tight time
  synchronization for wireless sensor networks,'' \emph{ACM Transactions on
  Sensor Networks (TOSN)}, vol.~3, no.~2, p.~8, 2007.

\bibitem{980173}
H.~Dai and R.~Han, ``Tsync: a lightweight bidirectional time synchronization
  service for wireless sensor networks,'' \emph{ACM SIGMOBILE Mobile Computing
  and Communications Review}, vol.~8, no.~1, pp. 125--139, 2004.

\bibitem{flashmob_portalacm}
T.~Baumgartner, S.~Fekete, W.~Hellmann, and A.~Kr\"{o}ller, ``Flash mob
  organization in heterogeneous wireless sensor networks,'' in \emph{NTMS'09:
  Proceedings of the 3rd international conference on New technologies, mobility
  and security}.\hskip 1em plus 0.5em minus 0.4em\relax Piscataway, NJ, USA:
  IEEE Press, 2009, pp. 487--490.

\bibitem{atmega48}
\BIBentryALTinterwordspacing
{Atmel Corporation}, ``Atmega48/88/168,'' 2009. [Online]. Available:
  \url{\url{http://www.atmel.com/dyn/resources/prod{\USCORE}
  documents/doc2545.pdf}}
\BIBentrySTDinterwordspacing

\bibitem{buschmann07isense}
\BIBentryALTinterwordspacing
C.~Buschmann and D.~Pfisterer, ``{iSense}: A modular hardware and software
  platform for wireless sensor networks,'' 6. Fachgespr{\"a}ch Drahtlose
  Sensornetze der GI/ITG-Fachgruppe Kommunikation und Verteilte Systeme, Tech.
  Rep., 2007. [Online]. Available:
  \url{\url{http://ds.informatik.rwth-aachen.de/events/fgsn07}}
\BIBentrySTDinterwordspacing

\bibitem{jennic}
\BIBentryALTinterwordspacing
{Jennic Ltd.}, ``{Product Brief JN513x, IEEE802.15.4 and ZigBee Wireless
  Microcontrollers},'' 2007. [Online]. Available:
  \url{\url{http://www.jennic.com/files/product{\USCORE}briefs/JN513x{\USCORE}%
PB{\USCORE} 160707{\USCORE}v1.2.pdf}}
\BIBentrySTDinterwordspacing

\end{thebibliography}

{\small
\vspace{3em}
\noindent

{\bf S{\'a}ndor P. Fekete} received his Diploma in Mathematics from
the University of Cologne, Germany (1989), and his Doctoral Degree in
Combinatorics and Optimization from the University of Waterloo
Science, Canada (1992). After being a postdoc at SUNY Stony Brook
(USA), he joined the Center for Parallel Computing in Germany in 1993
as an Assistant Professor, receiving his Habilitation in 1998. He was
an Associate Professor at the TU Berlin in 1999-2001, and joined the
Department of Optimization of the Braunschweig University of
Technology in 2001.

In 2007 he became chair for Algorithmics in the Department of Computer
Science. His research interests lie in both theoretical and practical
aspects of algorithms, with a large variety of interdisciplinary
collaborations. One of his main interests have been distributed
algorithms, in particular in the context of large sensor networks.

\vspace{3em}
\noindent

{\bf Alexander Kr{\"o}ller} received his Diploma in Mathematics from
TU Berlin, Germany (2003), and his Doctoral Degree in Mathematics from
Braunschweig University of Technology, Germany (2007).

He is currently an Academic Councilor at Braunschweig University of
Technology, Germany. He was a postdoc at SUNY Stony Brook, USA
(2008). He specializes in wireless sensor networks with a focus on
distributed algorithms.

\vspace{3em}
\noindent

{\bf Tobias Baumgartner} received his Diploma in Computer Science at
Braunschweig University of Technology, Germany (2006).

He is currently a PhD student in the Algorithms Group of Prof. Fekete
at Braunschweig University of Technology, Germany. In 2007, he worked
as a Software Developer in the field of wireless sensor networks for
ScatterWeb GmbH, Berlin, Germany. His main research interests are
distributed algorithms for wireless sensor networks, and programming
paradigms for tiny embedded systems---with a particular focus on
sensor nodes.

\vspace{3em}
\noindent

{\bf Winfried Hellmann} is a Bachelor student of Computer Science at
Braunschweig University of Technology, Germany. His research interests
lie in distributed algorithms for wireless sensor networks, and
firmware development for embedded systems.

%{\bf XXX} was born in Saigon, Vietnam.  He received his PhD degree in
%computer science from the University of Washington in 1999, his MS
%degree in electrical engineering and computer science from the
%Massachusetts Institute of Technology in 1988, and his BS degree in
%electrical engineering and computer science from the University of
%California, Berkeley in 1986.

%He is currently an Assistant Professor of Computer Science at
%Rutgers, the State University of New Jersey, USA.  He was a Member of
%Technical Staff at AT\&T Bell Laboratories from 1986 to 1991.  His
%current research interests include the performance, availability, and
%security of distributed systems, ranging from cluster-based Internet
%services to peer-to-peer systems.

%Prof. Nguyen is a member of the IEEE Computer Society and the ACM.
%He is a recipient of the US National Science Foundation CAREER award.

% The following kept the last paragraph above to be stretched
% spacing-wise.  I don't know why.
%~\\
}

\end{document}